\documentclass[%
reprint,
nofootinbib,
 amsmath,
 amssymb,
 aps,
pra,
]{revtex4-2}

\usepackage{graphicx}
\usepackage{dcolumn}
\usepackage{bm}
\usepackage{url}

\begin{document}


\title{Revisiting neutron propagation-based phase contrast imaging and tomography: use of phase retrieval to amplify the effective degree of brilliance}

\author{David M. Paganin}
\email{David.Paganin@monash.edu}
\affiliation{School of Physics and Astronomy, Monash University, Australia}

\author{Morten Sales}
\affiliation{Department of Physics, Technical University of Denmark, DK-2800 Kgs.~Lyngby, Denmark}

\author{Peter M. Kadletz}
\affiliation{European Spallation Source ESS ERIC, Lund, Sweden}

\author{Winfried Kockelmann}
\affiliation{STFC-Rutherford Appleton Laboratory, ISIS Facility, Harwell OX11 0QX, United Kingdom}

\author{Mario A. Beltran, Henning F. Poulsen, S{\o}ren Schmidt}
\affiliation{Department of Physics, Technical University of Denmark, DK-2800 Kgs.~Lyngby, Denmark~}





\date{\today}

\begin{abstract}

Propagation-based neutron phase-contrast tomography was demonstrated using the ISIS pulsed spallation source. The proof-of-concept tomogram with Paganin-type phase-retrieval filter applied exhibited an effective net boost of $23\pm 1$ in the signal-to-noise ratio as compared to an attenuation-based tomogram, implying a boost in the effective degree of neutron brilliance of over two orders of magnitude. This comparison is for phase retrieval versus conventional absorption with no additional collimation in place.  Expressions are provided for the optimal phase-contrast geometry as well as conditions for the validity of the method. The underpinning theory is derived under the assumption of the sample being composed of a single material.  The effective boost in brilliance may be employed to give reduced acquisition time, or may instead be used to keep exposure times fixed while improving the measured contrast.

\end{abstract}

\maketitle


\section{Introduction}

Propagation-based phase contrast has been known for millennia, e.g.~in its manifestation as heat shimmer over hot sand or in the twinkling of starlight.  The term ``propagation-based phase contrast'' arises due to the conversion of spatially-varying phase shifts in an optical wave field, namely spatial deformations in its associated wave fronts, into corresponding spatial intensity variations upon propagation \cite{Wilkins2014}.  
In a more modern setting, one has the out-of-focus contrast that is well known in visible-light microscopy \cite{Zernike1942,Bremmer1952} and electron microscopy \cite{CowleyBook}, together with propagation-based phase contrast for x rays \cite{Snigirev1995,Cloetens1996,Wilkins2014} and neutrons \cite{KleinOpat1975,KleinOpat1976,Allman2000}.

\citeauthor{KleinOpat1976}~\cite{KleinOpat1975,KleinOpat1976} gave an early demonstration of neutron propagation-based phase contrast, via Fresnel diffraction from a ferromagnetic domain in the context of demonstrating the spinor nature of neutrons. Later examples of neutron propagation-based phase contrast include \citeauthor{Allman2000}~\cite{Allman2000}, who also incorporated phase--amplitude retrieval using the algorithm of Paganin and Nugent \cite{paganin1998}.  For a textbook account of neutron phase contrast, see e.g. \citeauthor{NeutronOpticsHandbook} \cite{NeutronOpticsHandbook}, and for a recent review see e.g.~\citeauthor{Nelson2018} \cite{Nelson2018}.

In addition to propagation-based methods for achieving and subsequently decoding neutron phase contrast, which form the main focus of this paper, several well-developed methods must be mentioned.  Each has their relative strengths, which are clearly delineated in the series of contributions to the book edited by \citeauthor{NeutronImaghingAndItsApplications}~\cite{NeutronImaghingAndItsApplications}.  For the use of a two-crystal setup to achieve neutron phase contrast, in which the first crystal serves as a monochromator and the second as an analyzer, see e.g.~\citeauthor{Treimer2003}~\cite{Treimer2003}.  For the use of a Bonse--Hart \cite{BonseHart1965} three-blade interferometer cut from a single monolithic crystal, see e.g.~\citeauthor{Dubus2005}~\cite{Dubus2005}.   For grating-based neutron imaging, see \citeauthor{Pfeiffer2006}~\cite{Pfeiffer2006}.  This last-mentioned method has also been used to recover ultra-small-angle-scattering contrast (also spoken of as ``dark-field contrast'' within the neutron-imaging and x-ray-imaging communities\footnote{Note that this particular usage of the term ``dark field'' differs from the more general historical usage of the term, which has been employed for well over a century in the context of visible-light optics.  See e.g.~the 1920 review by Gage \cite{Gage1920}.}), that is due to unresolved micro-structure within the sample: see the paper on dark-field neutron tomography by \citeauthor{Strobl2008}~\cite{Strobl2008}. For summaries of all methods for neutron phase contrast, together with their relative strengths and differences, see e.g.~\citeauthor{Pfeiffer2009}~\cite{Pfeiffer2009} and \citeauthor{Kardjilov2018}~\cite{Kardjilov2018}.  

While all of the listed established methods for neutron phase contrast are of high importance, we restrict consideration to the propagation-based method for the remainder of the paper.  The key advantage we wish to seek is the possibility for boosting the effective (i.e., net) degree of brilliance associated with the neutron source.  This possible advantage is bought at the price of a significantly reduced degree of applicability compared to the other previously-cited methods, due to the strong enabling assumption that the sample is composed of a single material.  Moreover, the method of our paper is not able to access the dark-field signal mentioned in the previous paragraph.

One challenge of neutron phase contrast imaging in particular, and neutron imaging more generally, is low neutron brilliance.  Exposure times on the order of one hundred seconds are not unusual for neutron microscopy \cite{NeutronOpticsHandbook}, with exposures on the order of tens of seconds per projection in neutron tomography being typical (see e.g. \citeauthor{Zhang2018} \cite{Zhang2018}).  The low brilliance often requires one to highly collimate the neutron beam to observe coherence-based effects such as propagation-based phase contrast, further increasing the necessary exposure times.  All of the above leads to typical neutron-tomography exposure times on the order of hours (see e.g.~\citeauthor{LaManna2017} \cite{LaManna2017}), with total tomographic scan times on the order of 10~s being deemed ``ultra fast'' and two-dimensional image-acquisition times on the order of a second being described as having ``high temporal resolution'' \cite{Totzske2017,Totzke2019}.  Very recently, fast neutron tomography using exposure times, for each orientation of a sample, of 0.2s has been reported, in which 600 projections were obtained in 2 minutes \cite{Totzske2021}.  

The possibility for further reductions in such acquisition times, for neutron imaging in both two and three spatial dimensions, is a key driver for this paper.  Rather than seeking an increase in the physical neutron source brilliance or detector efficiency, we instead consider how propagation-based phase contrast and subsequent phase-retrieval decoding may lead to an increase in the effective degree of brilliance for existing sources.  This is achieved in the following indirect manner, the first of which degrades the pre-phase-retrieval physical degree of brilliance and the second of which enhances the post-phase-retrieval degree of effective (i.e.,~net) brilliance: 
\begin{enumerate}
  \item We first sufficiently collimate the neutron beam in order to have sufficient spatial coherence for propagation-based phase contrast to be manifest \cite{KleinOpat1975,KleinOpat1976,Allman2000}. 
  \item Subsequent application of a phase retrieval step \cite{paganin2002}, to the propagation-based phase contrast images, boosts the effective signal-to-noise ratio (SNR) of the processed image.
\end{enumerate}

The above two-step procedure will only be useful for scenarios in which there is a net increase in the effective degree of brilliance, namely if the second step's implied increase in the effective degree of brilliance outweighs the first step's reduction in actual brilliance.  This criterion specifies the domain of practicality of the method.  It must be emphasized that this SNR comparison is for the case of phase retrieval versus conventional absorption with no additional collimation in place.

The above strategy may be viewed as a special case of an extremely general approach to signal transmission that is very well known in information theory \cite{MacKay2003}.  Under this view, one may transmit information through a noisy information channel using the following process:
\begin{enumerate}
\item Suitably encode the signal that is to be transmitted.
\item Transmit the encoded signal through a noisy communication channel.
\item Decode the received signal.
\end{enumerate}
  The key point, behind this indirect information-transmission paradigm, is that a suitable choice for the encoding may enable accurate transmission of information through noisy channels.  In the present context of propagation-based neutron imaging and tomography, information regarding a sample is imprinted upon a neutron beam that traverses that sample.  Subsequent propagation-based phase contrast constitutes a form of coding, of the object's transmission function, that is induced by the diffraction physics of free-space neutron propagation between the sample and the position-sensitive detector.  The phase-retrieval step \cite{paganin2002}, which may be applied to the measured image, may be viewed as a decoder in the information-transmission channel associated with the process of phase contrast imaging.  For the admittedly restrictive class of samples that may be approximately modeled as composed of a single material of possibly-varying density, and under the additional approximations that both the effective source size and the object-to-detector distance are sufficiently small, the phase-retrieval decoding yields a unique reconstruction for the sample's transmission function \cite{paganin2002}. While the strong limitation to single-material samples is an important and significant restriction on the domain of applicability of the neutron phase-contrast method reported here, a range of samples of applied-physics importance meet this requirement.  

 We close this introduction with a summary of the remainder of the paper.  Section~\ref{Sec:Theory} outlines how propagation-based neutron phase contrast, for the special case of a single-material sample, may be employed for significant boosts in the effective degree of  brilliance of sources used for neutron imaging.  This boost, which may be traded off against significant reductions in acquisition time, is enabled by the code--decode paradigm associated with the constructive use of propagation-based phase contrast.  We give a simple experimental demonstration establishing proof-of-concept for these ideas, for neutron imaging in both two and three dimensions, in Sec.~\ref{Sec:Experiment}.  Some broader implications of this work are discussed in Sec.~\ref{Sec:Discussion}, together with some suggested avenues for future work.  We conclude with Sec.~\ref{Sec:Conclusion}. 

\section{Theory of effective-brilliance amplification for neutron imaging}\label{Sec:Theory}

\subsection{Qualitative explanation of the method}

Before presenting the theory underpinning the method, we describe its key aspects in qualitative terms.  

\begin{figure*}
\includegraphics[width=0.95\textwidth,scale=0.09]{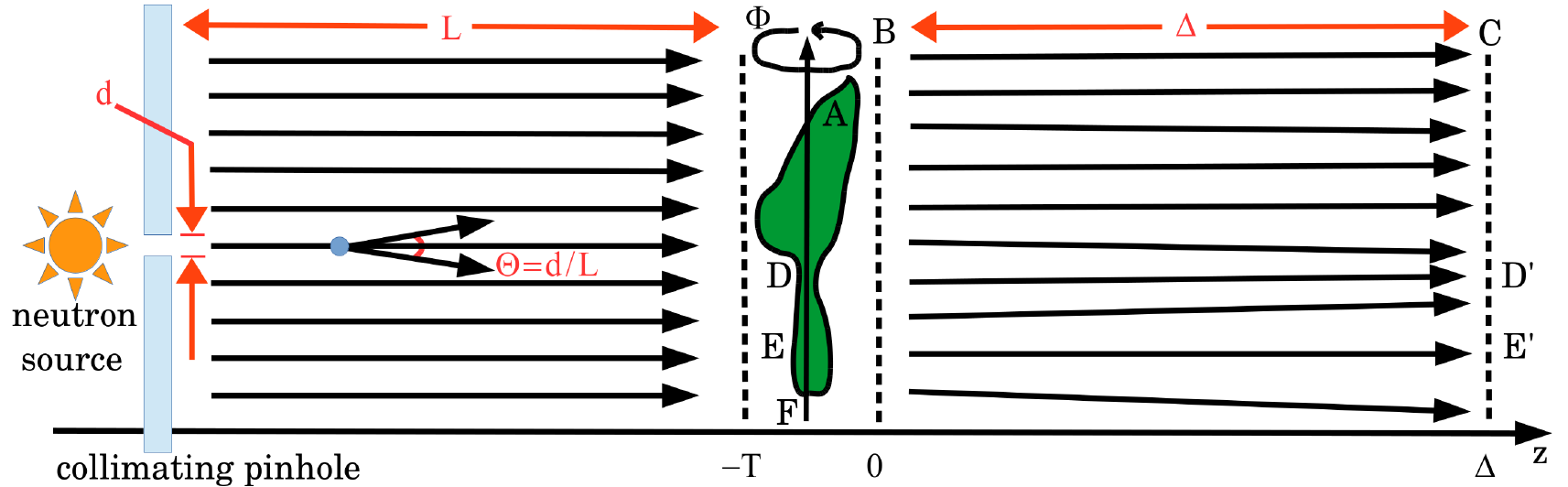}
\caption{Generic setup for effective-brilliance amplification in neutron phase contrast imaging and tomography.  Neutrons are collimated using a pinhole with diameter $d$, before propagating a distance $L$ so as to illuminate a thin single-material object $A$ with number density of nuclei $\rho(x,y,z)$ that is confined to the region $-T<z<0$.  The transmitted neutrons propagate from the exit surface $B$ to a pixellated energy-resolving detector $C$, at distance $\Delta$ downstream of $B$.}
\label{fig:SetUp}
\end{figure*}

Consider Fig.~\ref{fig:SetUp}, which shows a static non-magnetic sample $A$, under the strong assumption that the said sample is composed of a single material.  The density of this material is allowed to be variable. The sample is, by assumption, sufficiently weakly attenuating that a contact image, registered over the plane $B$ at the exit surface of the sample, will display little contrast.  

However if this sample is illuminated by a parallel neutron beam that has been sufficiently highly collimated, the angular divergence $\Theta$ of the neutron beam will be sufficiently small for the refractive properties of the sample to be visible when a propagated image is measured over the plane $C$ at some distance $\Delta$ downstream of the sample. More precisely, we need the penumbral blur width
\begin{equation}
 W=\Theta\Delta    
\end{equation}
over the plane $z=\Delta$, to not be so large as to wash out the subtle refractive features due to the sample, in the image that is registered over the plane $z=\Delta$.  

These refractive features are analogous to the pattern of bright lines seen at the bottom of a swimming pool on a hot sunny day.  Rather than the surface of the water in the pool serving to refract sunlight, here local spatial variations in the refractive index of the sample serve to refract neutrons.  

Thus e.g.~if we restrict ourselves to sample materials having a refractive index that is less than unity, as is often but not always the case (see e.g.~Table 8.1 in \citeauthor{Pfeiffer2009} \cite{Pfeiffer2009} for some exceptions), the concave feature at $D$ will act as a converging neutron lens with focal length that is much larger than $\Delta$. Hence the neutron intensity measured at $D'$ will be slightly larger due to the converging effect of neutrons passing through point $D$ in the sample. Similarly, convex points on the sample such as $E$ act as locally-diverging neutron lenses, causing the intensity at $E'$ to be slightly less than it would have been in the absence of the sample.  

This mechanism, using free-space propagation to convert the refractive effects of a sample into contrast in a measured neutron image \cite{KleinOpat1976,Allman2000}, is known as propagation-based phase contrast.  This particular use of the term ``phase'' refers to the phase of the neutron wavefunction $\Psi$, since surfaces of constant phase
\begin{equation}
\varphi\equiv\arg\Psi=~\textrm{constant}
\end{equation}
define neutron wave fronts.  The refractive effects we have described may be considered as arising from distortions in the initially-planar neutron wave front arising from its passage through the sample {\em en route} to the detector plane $z=\Delta$.  

One key aspect, of this form of contrast, is that it emphasizes very fine features that are present in the sample. This image-sharpening behavior is a consequence of the fact that ``a smaller lens is a stronger lens'', thus e.g.~the smaller the radius of curvature of the concavity at $D$ or the convexity at $E$ in Fig.~\ref{fig:SetUp}, the greater the consequent intensity increase at $D'$, and the greater the consequent intensity decrease at $E'$.  Again, we note that in order to see this effect, the degree of collimation of the neutron beam must be sufficiently high that the penumbral-blur width $\Theta\Delta$ is smaller than the transverse length scale of the fine-level spatial structures that are amplified in visibility in the process of propagation-based neutron phase contrast.  Stated more succinctly, the degree of collimation must be sufficiently large and the sample-to-detector distance must be sufficiently small (but not too small). 

Propagation-based neutron phase contrast images, unlike their attenuation-contrast counterparts, bear only an indirect relationship to the sample being imaged.  This is because a phase-retrieval decoding step is required, in order to obtain quantitative information regarding a sample.  Hence the inverse problem \cite{Sabatier1990} of processing the measured propagation-based phase contrast image over the plane $z=\Delta$, to give the projected density of the sample.  Since features in the sample are sharpened upon propagation from the exit-surface of the object to the surface of the detector, the measured fine spatial detail in the image will need to be blurred (low-pass filtered) in a suitable manner that is consistent with the diffraction physics of phase-contrast image formation, in the process of decoding the measured intensity so as to obtain the projected density of the sample.  

A means for doing so is outlined in the next subsection, leading to the five-step algorithm at the end of Sec.~II.B. As shall be seen in due course, this indirect procedure for imaging neutron-transparent single-material samples has the attractive attribute that it boosts the effective degree of brilliance of the neutron source.  Thus we trade off (i) the noise-increasing results of collimating the divergence of the neutron beam $\Theta$ to a smaller value than one would otherwise use, and/or reducing the acquisition time, against (ii) the noise-suppressing effect of the effective-brilliance boost, associated with the subsequent phase-retrieval data processing.  

Impetus is given to follow the indirect procedure explored in this paper, by the fact that the method---namely the so-called Paganin method, which originated in a 2002 paper applied to x rays \cite{paganin2002}---can boost the SNR of reconstructed propagation-based phase contrast images by over two orders of magnitude for both x rays  \cite{SNRboost1,SNRboost2,SNRboost3,SNRboost4,SNRboost5} and electrons \cite{Clark2019}.  This SNR boost may be used to decrease x-ray acquisition time by over four orders of magnitude.  For x rays, the noise-suppression property of the Paganin phase-retrieval method has been used to enable significantly reduced acquisition time, to the extent where over 200 \cite{200tps} and even 1000 \cite{1000tps} x-ray phase-contrast tomograms per second is now possible.  Alternatively, acquisition time can be kept fixed to a usual duration (i.e.,~tens of seconds per projection for neutron tomography), in which case the application of the method will enable increased 
contrast.   

\subsection{Mathematical basis of the method}

Figure~\ref{fig:SetUp} shows a static non-magnetic sample $A$ composed of a single material.  Both the number density of nuclei and the number density of atoms are assumed to be equal to one another, being denoted by $\rho({\bf r}_{\perp},z)$, where ${\bf r}_{\perp}=(x,y)$ denote transverse spatial coordinates in planes perpendicular to the optical axis $z$.  Neutrons propagating in the $z$ direction are transmitted through the sample, before traveling a distance $\Delta$ and having their resulting spatial intensity distribution $I({\bf r}_{\perp},z=\Delta)$ recorded by a pixellated planar energy-resolving detector $C$. Mono-energetic neutrons are assumed here, and for the subsequent experiment, with Appendix A suggesting how this theory might be generalized to the case of broad-band poly-energetic illumination. 

The neutron radiation is considered to have a divergence \cite{Treimer2009}
\begin{equation}\label{eq:DivergenceDefined}
\Theta=d/L, 
\end{equation}
which is sufficiently small that the associated penumbral blur width $\Theta\Delta$, over the plane $z=\Delta$, is less than the transverse shift of neutrons, in the plane $z=\Delta$, that is induced by the refractive effects of the sample. Here, $d$ is the diameter of the collimating pinhole, and $L$ is the distance from the collimating pinhole to the sample.

%
%

Assume that the degree of collimation is  sufficiently high that the image, measured over the plane $z=\Delta$, indeed exhibits propagation-based neutron phase contrast (see next sub-section for further detail on this criterion).  We can then turn attention to the corresponding inverse problem of recovering the object projected density, given the propagation-based phase contrast image $I({\bf r}_{\perp},z=\Delta)$ as input data.  If the object is rotated step-wise around the axis $F$ through a variety of angles $\Phi$ (see Fig.~\ref{fig:SetUp}), and an image over plane $C$ recorded for each $\Phi$ step, the projected density for each angular orientation of the object may subsequently be used for a tomographic reconstruction of the density \cite{GureyevAPL2006}.     

The sample's complex refractive index is \cite{CremerChap6}:
\begin{align}\label{eq:ComplexRefractiveIndex}
n({\bf r}_{\perp},z)=1-\delta({\bf r}_{\perp},z)+\frac{i\lambda}{4\pi}\mu({\bf r}_{\perp},z),
\end{align}
where $\lambda$ is the wavelength and $\mu$ is the linear attenuation coefficient. Here, the previously mentioned strong assumption of a single-material sample enables us to use the Fermi thin-slab formula \cite{Sears1982, Warner1985}
\begin{align}\label{eq:Expression_for_delta}
\delta({\bf r}_{\perp},z)=b \rho({\bf r}_{\perp},z) \lambda^2/(2\pi)
\end{align}
for the refractive index decrement, and 
\begin{align}\label{eq:Expression_for_mu}
\mu({\bf r}_{\perp},z)=\sigma\rho({\bf r}_{\perp},z) 
\end{align}
for the linear attenuation coefficient.  Here, $b$ is the bound neutron scattering length and $\sigma$ is the total neutron cross section\footnote{Values for the neutron scattering lengths and cross sections are tabulated at the National Institute of Standards and Technology (NIST) Center for Neutron Research website at \url{https://ncnr.nist.gov/resources/n-lengths/}.}.

The projection approximation \cite{paganin2006} is assumed to hold, hence the intensity $I$ and phase $\varphi$ of the neutron beam, at a specified energy $E$ over plane $B$, are given by:
\begin{align}
I({\bf r}_{\perp},z=0) &= I_0 \exp\left[-\sigma\rho_{\perp}({\bf r}_{\perp})\right], \label{eq:IntensityProjectionApproximation}\\
\varphi({\bf r}_{\perp},z=0) &= -b \, \lambda \, \rho_{\perp}({\bf r}_{\perp}).\label{eq:PhaseProjectionApproximation}
\end{align}
Here, $I_0$ is the intensity of the illuminating neutron beam, $\rho({\bf r}_{\perp},z)$ is assumed to only be non-zero in the volume $-T\le z\le 0$, and $\rho_{\perp}({\bf r}_{\perp})$ is the projected density:
\begin{align}
\rho_{\perp}({\bf r}_{\perp})\equiv\int_{-T}^{0} \rho({\bf r}_{\perp},z)dz.
\end{align}

Assuming all streamlines of the neutron current density in the slab $-T \le z \le 0$ to be almost parallel with the $z$ axis, we can invoke the paraxial form of the continuity equation implied by the time-independent Klein--Gordon equation.  Known as the transport-of-intensity equation \cite{teague1983} in a phase-retrieval context, this paraxial time-independent special case of Madelung's ``hydrodynamical'' formulation of quantum mechanics \cite{Madelung1926} is:
\begin{align}\label{eq:TIE}
-\frac{\lambda}{2 \pi}\nabla_{\perp}\cdot{[I({\bf r}_{\perp},z=0)\nabla_{\perp}\varphi({\bf r}_{\perp},z=0)]}=\left. \frac{\partial I({\bf r}_{\perp},z)}{\partial z} \right  \rvert_{z=0}.
\end{align}
Here, $\nabla_{\perp}\equiv(\partial/\partial x,\partial/\partial y)$ denotes the gradient operator in the $xy$ plane.  Note that the term in square brackets, above, is proportional to the transverse component of the flow vector associated with the paraxial neutron propagation.   Given that the negative divergence of a paraxial vector field is a measure of its rate of convergence, we see that the above paraxial continuity equation makes the intuitively-reasonable statement that the convergence of the neutron flow is proportional to the longitudinal rate of change of the corresponding intensity.  

Let us further assume the object-to-detector distance $\Delta$ to be sufficiently small that the resulting Fresnel diffraction pattern at $z=\Delta$ is in the near field of the object.  We may now adapt the logic in \citeauthor{paganin2002} \cite{paganin2002} from x rays to neutrons, as follows:  (i) Make a first-order finite-difference approximation to the right-hand side of Eq.~(\ref{eq:TIE}), using the intensity distributions over the planes $z=0$ and $z=\Delta$; (ii) use Eqs.~(\ref{eq:IntensityProjectionApproximation}) and (\ref{eq:PhaseProjectionApproximation}); (iii) isolate the intensity over the plane $z=\Delta$ in the resulting expression. This gives (cf. Eq.~(7) in \citeauthor{paganin2002} \cite{paganin2002}):
\begin{align}\label{eq:TIE-Hom}
\frac{I({\bf r}_{\perp},z=\Delta)}{I_0}=\left( 1-\frac{b\lambda^2\Delta}{2\pi\sigma}\nabla_{\perp}^2\right) \exp\left[-\sigma\rho_{\perp}({\bf r}_{\perp})\right],
\end{align}
where $\nabla_{\perp}^2$ denotes the Laplacian in the $xy$ plane.  The propagation-based phase contrast in the recorded intensity distribution $I({\bf r}_{\perp},z=\Delta)$ is manifest in the Laplacian term on the right side, with its multiplier proportional to the object-to-detector distance $\Delta$. 

The right-hand side of Eq.~(\ref{eq:TIE-Hom}) is mathematically identical to the application of a Laplacian-type unsharp-mask image sharpening operator \cite{Unsharp0,Gureyev2004,Unsharp1,Unsharp2} to the attenuation-contrast image $I_0 \exp\left[-\sigma\rho_{\perp}({\bf r}_{\perp})\right]$, an observation which renders precise our earlier qualitative statements that propagation-based phase contrast serves to sharpen an image.  As we show later, this observation is key to the boost in the effective degree of brilliance that is inherent in the method considered in this paper.    

To proceed further, Fourier transform Eq.~(\ref{eq:TIE-Hom}) with respect to $x$ and $y$, utilize the Fourier derivative theorem, solve the resulting algebraic equation for the Fourier transform of $\exp[-\sigma\rho_{\perp}({\bf r}_{\perp})]$, then inverse Fourier transform and solve for the projected density $\rho_{\perp}({\bf r}_{\perp})$.  This gives the neutron-optics form of a phase-retrieval algorithm previously published for x rays \cite{paganin2002}:
\begin{align}\label{eq:PaganinMethodForNeutrons}
\rho_{\perp}({\bf r}_{\perp})=-\frac{1}{\sigma}\log_{\textnormal{e}}\left(\mathcal{F}^{-1}\left\{\frac{\mathcal{F}[I({\bf r}_{\perp},z=\Delta)/I_0]}{1+\frac{b\lambda^2\Delta}{2\pi\sigma}(k_x^2+k_y^2)}\right\}\right).
\end{align}
Here, $(k_x,k_y)$ are Fourier-space spatial frequencies corresponding to $(x,y)$, $\mathcal{F}$ denotes Fourier transformation with respect to $x$ and $y$, and $\mathcal{F}^{-1}$ is the corresponding inverse transform with respect to $k_x$ and $k_y$. We have used a convention in which Fourier transformation converts $\partial/\partial x$ to $ik_x$ and $\partial/\partial y$ to $ik_y$, according to the Fourier derivative theorem.  Note that we require $b \ge 0$ to avoid a division-by-zero singularity in the denominator of Eq.~(\ref{eq:PaganinMethodForNeutrons}). This is indeed the case for most but not all materials at thermal and cold neutron energies.  Manganese, titanium, vanadium, lithium, and the $^1$H isotope of hydrogen (together with certain compounds thereof) are important exceptions \cite{Pfeiffer2009}.\footnote{Actually, the more general requirement is that $b \Delta$ be greater than or equal to zero.  Thus, if $b$ is positive then $\Delta$ should also be positive, which corresponds to the scenario that is considered in the present paper.  Alternatively, if $b$ is negative, then $\Delta$ should be negative.  Such a regime, of negative $\Delta$, can be achieved using negative defocus if a lens-based neutron imaging system (which is capable of both under-focus and over-focus) is used to form the registered neutron phase contrast image. }

Since blurring due to finite source size is important for typical neutron sources, we follow \citeauthor{SNRboost1} \cite{SNRboost1} in noting that a stable partial deconvolution for the effective source-blurring area 
\begin{equation}\label{eq:DivergenceRelatedToBlurArea}
 \mathcal{A}=W^2=(\Theta\Delta)^2   
\end{equation}
(referred to the imaging plane $z=\Delta$) can be achieved using the following logic.  Blurring of a two-dimensional image, over the transverse length scale $W$, can be achieved by applying the blurring operator $1+\frac{1}{8}W^2\nabla_{\perp}^2$ to that image \cite{Unsharp0,Gureyev2004,beltran2018}.  Hence the effects of divergence-induced blurring may be taken into account by acting on the right side of Eq.~(\ref{eq:TIE-Hom}) with the operator
\begin{equation}
1+\frac{1}{8}W^2\nabla_{\perp}^2\approx 1+\frac{1}{8}\mathcal{A}\nabla_{\perp}^2, 
\end{equation}
to give an equation of the Fokker--Planck type \cite{Risken1989,beltran2018,MorganPaganin2019,PaganinMorgan2019}:
\begin{align}\label{eq:TIE-HomWithBlurring}
\frac{I({\bf r}_{\perp},z =\Delta)}{I_0} \!\!\!\!\!\!\!\!\!\!\!\!\!\!\!\!\!  \\  \nonumber &=\left[ 1+\frac{1}{8}\mathcal{A}\nabla_{\perp}^2\right] \left[ 1-\frac{b\lambda^2\Delta}{2\pi\sigma}\nabla_{\perp}^2\right] \exp\left[-\sigma\rho_{\perp}({\bf r}_{\perp})\right] \\ \nonumber &\approx \left[ 1-\left(\frac{b\lambda^2\Delta}{2\pi\sigma}-\frac{\mathcal{A}}{8}\right)\nabla_{\perp}^2\right] \exp\left[-\sigma\rho_{\perp}({\bf r}_{\perp})\right].
\end{align}
Note that a term, containing the bi-Laplacian operator $\nabla_{\perp}^2\nabla_{\perp}^2$, has been discarded in the last line of the above equation.  Comparison of Eq.~(\ref{eq:TIE-Hom}) with Eq.~(\ref{eq:TIE-HomWithBlurring}) shows that, to account for divergence-induced blurring, we need to make the following replacement in Eq.~(\ref{eq:PaganinMethodForNeutrons})~\cite{beltran2018}:
\begin{equation}\label{eq:ReplacementDueToBlurring}
    \frac{b\lambda^2\Delta}{2\pi\sigma} \longrightarrow \frac{b\lambda^2\Delta}{2\pi\sigma} - \frac{\mathcal{A}}{8}.
\end{equation}
This replacement extends Eq.~(\ref{eq:PaganinMethodForNeutrons}) into the main result of this paper (cf.~Eq.~(8) of \citeauthor {beltran2018} \cite{beltran2018}):
\begin{align}\label{eq:PaganinBeltranMethodForNeutrons}
\rho_{\perp}({\bf r}_{\perp})=   -\frac{1}{\sigma}\log_{\textnormal{e}}\left(\mathcal{F}^{-1}\left\{\frac{\mathcal{F}[I({\bf r}_{\perp},z=\Delta)/I_0]}{1+\tau(k_x^2+k_y^2)}\right\}\right),
\end{align}
where
\begin{align}\label{eq:tau}
\tau=\frac{\lambda^2 b \Delta}{2\pi\sigma}-\frac{\mathcal{A}}{8}=\frac{1}{2}\left[\frac{\lambda^2 b \Delta}{\pi\sigma}-\frac{(\Theta\Delta)^2}{4}\right].
\end{align}
The above expression permits the projected density $\rho_{\perp}({\bf r}_{\perp})$ of the single-material sample, to be uniquely obtained from a single propagation-based neutron phase contrast image $I({\bf r}_{\perp},z=\Delta)$.  With the exception of the numerically trivial multiplicative constant $1/\sigma$, this decoding of the propagation-based phase contrast image depends on the single parameter $\tau > 0$. The core of the algorithm is thus rather simple, being the low-pass Lorentzian Fourier-space filter $1/[1+\tau(k_x^2+k_y^2)]$.

The analysis process, given in Eqs.~(\ref{eq:PaganinBeltranMethodForNeutrons}) and (\ref{eq:tau}), is equivalent to the following computationally-simple algorithm:

\begin{enumerate}
    \item Take a single propagation-based phase contrast  neutron image with measured intensity distribution $I({\bf r}_{\perp},z=\Delta)$, as a function of 2D position coordinates ${\bf r}_{\perp}$ in the detector plane, and then normalize (or, more generally, flat-field correct) via division by the background intensity $I_0$.
    \item Apply a two-dimensional fast Fourier transform $\mathcal{F}$ to the normalized image, thereby generating a complex image that is a function of the Fourier-space coordinates $(k_x,k_y)$.  Each pixel in this Fourier-space image will have height (width) equal to $1/W_x$ or $1/W_y$, where $W_x~(W_y)$ is the physical width (height) of the original image input into Step \#1 above.
    \item Multiply the result of Step \#2 by the low-pass Fourier filter (Lorentzian filter) $1/[1+\tau(k_x^2+k_y^2)]$, where the numerical value of $\tau$ is given by Eq.~(\ref{eq:tau}).  Optional: As an alternative to using Eq.~(\ref{eq:tau}) to calculate $\tau$, for non-quantitative studies we may simply tune this parameter according to the criterion that $\tau$ should be sufficiently small to eliminate phase-contrast fringes from the data, but not so large as to introduce excessive blurring.  
    \item Apply an inverse fast Fourier transform $\mathcal{F}^{-1}$.
    \item Take the natural logarithm of the resulting image, and then divide by $-\sigma$.  The resulting image is a map of the projected density $\rho_{\perp}({\bf r}_{\perp})$.  For non-quantitative studies, such as the experimental study presented in our paper,    the division by $-\sigma$ may be omitted.
\end{enumerate}

In a tomographic setting, the above steps may be applied to each projection, with the resulting processed images subsequently being input into a conventional tomography reconstruction procedure such as filtered-backprojection or algebraic reconstruction methods \cite{KakSlaneyBook}.  

\subsection{Choice of degree of neutron collimation}\label{Sec:ChoiceOfCollimation}

A condition that the denominator in Eq.~(\ref{eq:PaganinBeltranMethodForNeutrons}) never vanishes is that $\tau > 0$.  Making use of the second definition for $\tau$ in Eq.~(\ref{eq:tau}), we obtain the collimation condition:
\begin{equation}\label{eq:CollimationCondition2}
    \Theta < \Theta_{\textrm{critical}}=2\lambda\sqrt{\frac{b}{\pi\sigma\Delta}}=\sqrt{\frac{8\delta}{\Delta\mu}}.
\end{equation}
This condition may be thought of as quantifying the need for sufficiently high spatial coherence, for the propagation-induced neutron phase contrast to be non-negligible \cite{Mishra2011}.  It may therefore be considered as a rule of thumb required to be in a phase-contrast imaging regime, which ensures that the penumbral blurring due to non-zero divergence does not entirely wash out propagation-based phase contrast (cf.~\citeauthor{Gureyev2008} \cite{Gureyev2008}).  Note that the various functional dependencies in Eq.~(\ref{eq:CollimationCondition2}) make intuitive physical sense.  Thus, the larger the object-to-detector distance $\Delta$, the greater the effect of penumbral blurring and hence the more stringent is the required collimation condition.  Conversely, the smaller the value of $b/\sigma$ or $\delta/\mu$, the weaker the propagation-based phase contrast signal and hence the more stringent the collimation condition needs to be.

The collimation condition in Eq.~(\ref{eq:CollimationCondition2}) immediately implies the following trade-off.  Greater beam divergence $\Theta$ gives more neutrons and therefore better statistics, but worse phase contrast on account of the associated blurring leading to suppressed propagation-based phase contrast.  Conversely, smaller beam divergence gives improved phase contrast on account of the improved spatial coherence, at the expense of higher noise in the measurement statistics.  What is an optimum degree of collimation, given this trade-off?

The term in large round brackets on the final line of Eq.~(\ref{eq:TIE-HomWithBlurring}), namely $\tau$ as given in Eq.~(\ref{eq:tau}), is a measure of the visibility of the propagation-based phase contrast features that we  would observe in the high-neutron-flux limit (zero noise limit).  See Appendix B for a justification of this claim.  This noise-free-case visibility decreases with increasing divergence $\Theta$, since $\tau$ gets smaller as $\Theta$ gets bigger.  Conversely, the SNR associated with a finite neutron flux, will scale proportionally to the square root of the solid angle $\Theta^2$ associated with the divergence, so that  SNR $\propto\Theta$.  Therefore the quantity we need to extremize with respect to $\Theta$ is $\tau \textrm{ SNR }$ (cf.~Rule \#3 in \citeauthor{Gureyev2008} \cite{Gureyev2008}):
\begin{equation}
 \tau \textrm{ SNR } \propto \left[\frac{\lambda^2 b \Delta}{\pi\sigma}-\frac{(\Theta\Delta)^2}{4} \right] \Theta.    
\end{equation}
Differentiating the above equation with respect to $\Theta$, and setting the result to zero, gives an optimal divergence $\Theta_{\textrm{optimum}}$ that is $1/{\sqrt{3}}\approx 60\%$ of the maximum value given in the collimation condition (Eq.~(\ref{eq:CollimationCondition2})).  Thus we have the optimum divergence
\begin{align}\label{eq:OptimumDivergence}
    \Theta_{\textrm{optimum}} = (1/\sqrt{3}) ~ \Theta_{\textrm{critical}}.
\end{align}

This gives rise to another simple rule of thumb: Reduce the divergence to approximately 60\% of the value at which phase contrast is first observed in the image, to give a near-optimal experiment in light of the trade-off mentioned in the previous paragraph.  Further convenient rules of thumb, for optimized propagation-based phase contrast imaging, can be found in the previously cited paper by \citeauthor{Gureyev2008}~\cite{Gureyev2008}.  Note also, that in line with the well-known trade-off between noise and resolution \cite{GureyevNRU}, one may opt to reduce the divergence below the ``optimum effective brilliance'' value given above.

\subsection{Amplification of the effective degree of brilliance}

The single-parameter reconstruction in Eq.~(\ref{eq:PaganinBeltranMethodForNeutrons}) is mathematically identical to the so-called Paganin method \cite{paganin2002}.  This latter algorithm, which has been applied in several hundred papers in an x-ray setting \cite{GPM2020}, exhibits extreme stability with respect to noise in the input phase-contrast image \cite{gureyev2017unreasonable}.  Indeed, for x rays the algorithm has been seen to boost SNR by two orders of magnitude or more, enabling acquisition times to be reduced by four orders of magnitude or more \cite{SNRboost1,SNRboost2,SNRboost3,SNRboost4,SNRboost5}.  

At this point we must emphasize that for neutrons the situation is less favorable---but nevertheless worth investigating, in our view---for two distinct reasons.  (i) Firstly, the broad domain of applicability of the single-material assumption \cite{GPM2020}, in an x-ray setting, arises from the fact that the Klein--Nishina formula of relativistic quantum mechanics \cite{KleinNishina} implies that low-atomic-number materials at sufficiently high x-ray energy will asymptote towards a single material, namely electrons, with their associated  electron density governing the manner in which the material interacts with x rays \cite{KleinNishina2}.  This simplification does not apply to the more complex case of neutron phase-contrast imaging, since no such asymptotic behavior is operative for neutron scattering.  (ii) For the case of neutron imaging, the previously mentioned SNR boost must be traded off against the SNR reduction that results from  collimating the divergence to a sufficient degree that the collimation condition in Eq.~(\ref{eq:CollimationCondition2}) is satisfied.  We consider this trade-off in more detail below, with a view to determining the conditions under which the loss in flux, due to increased collimation, is sufficiently compensated by the boost in the effective degree of brilliance that arises from the subsequent phase-retrieval decoding. 

Let $\Theta_0 > \Theta_{\textrm{critical}}$ denote the beam divergence that would be utilized in the context of a particular experiment utilizing attenuation-contrast neutron imaging.  If we collimate sufficiently strongly that the divergence is now reduced to $\Theta_{\textrm{critical}}$, the corresponding flux will scale by the multiplicative factor $(\Theta_{\textrm{critical}}/\Theta_0)^2$.  This leads to a loss in SNR corresponding to the multiplicative factor
\begin{equation}
    f = \frac{\Theta_{\textrm{critical}}}{\Theta_0}=\frac{2\lambda}{\Theta_0}\sqrt{\frac{b}{\pi \sigma \Delta}}.
\end{equation}
%

Next, we apply the formulae of \citet{SNRboost3} and \citet{SNRboost4}.  There, the maximum SNR gain (due to the phase-retrieval step) $G_{\textnormal{max}}$ in a tomographic setting\footnote{The maximum SNR gain is different for two-dimensional imaging (radiography) versus three-dimensional imaging (tomography).  The gain is somewhat larger for the tomographic case \cite{SNRboost3,SNRboost4}.} is shown to be  
\begin{align}
G_{\textnormal{max}}\approx 0.3 \frac{\delta}{\beta}\approx\frac{\pi\delta}{\lambda\mu},
\end{align}
where  
\begin{equation}
\beta=\textrm{Im}(n)=\frac{\lambda\mu}{4\pi},
\end{equation}
and we have made the approximation that $0.3\times 4 \approx 1$.  Using Eqs.~(\ref{eq:Expression_for_delta}) and (\ref{eq:Expression_for_mu}),
\begin{align}\label{eq:GainIntermediateExpression}
G_{\textnormal{max}}\approx \frac{b\lambda}{2\sigma}.
\end{align}

To take the effects of penumbral blurring into account, Eq.~(\ref{eq:ReplacementDueToBlurring}) implies that we must make the following replacement in Eq.~(\ref{eq:GainIntermediateExpression}):
\begin{equation}\label{eq:ReplacementDueToBlurring2}
    \frac{b\lambda}{2\sigma} \longrightarrow \frac{b\lambda}{2\sigma} - \frac{\mathcal{\pi A}}{8\lambda\Delta},
\end{equation}
leading to:
\begin{align}\label{eq:GainIntermediateExpression2}
G_{\textnormal{max}}\longrightarrow \frac{b\lambda}{2\sigma}- \frac{\mathcal{\pi A}}{8\lambda\Delta}.
\end{align}

The net corresponding boost in effective brilliance, when both the collimation SNR loss and the phase retrieval SNR boost are taken into account, is:
\begin{align}\label{eq:X}
B_{\textnormal{max}}=f^2 G_{\textnormal{max}}^2=\frac{b\lambda^2}{\pi\sigma\Delta\Theta_0^2}\left(\frac{b\lambda}{\sigma}-\frac{\pi\mathcal{A}}{4\lambda\Delta}\right)^2>1.
\end{align}

If we choose the optimal collimation given by Eq.~(\ref{eq:OptimumDivergence}), then we may take 
\begin{equation}
\mathcal{A}=\Theta_{\textrm{optimum}}^2\Delta^2 
\end{equation}
and so Eq.~(\ref{eq:X}) becomes:
\begin{align}\label{eq:Y}
B_{\textnormal{max}}=\frac{4 b^3 \lambda^4}{9\pi\sigma^3\Delta\Theta_0^2}=\frac{\lambda(\delta/\beta)^3}{18\pi\Delta\Theta_0^2}>1.
\end{align}

It is only when the above inequality is satisfied, i.e.,~when
\begin{equation}
B_{\textnormal{max}}>1, 
\end{equation}
that there is a net boost in the effective degree of brilliance on account of the combined effects of increased collimation followed by phase retrieval.  If the inequality in Eq.~(\ref{eq:Y}) is not satisfied, there is no net benefit to be obtained using a propagation-based single-material phase-contrast approach.  

We close this section by noting that the maximal phase-retrieval-enabled  SNR boost $B_{\textrm{max}}$ is strongly material dependent, as it is proportional to $(\delta/\beta)^3$. As an indicative example, consider the elements Pb, Si, Cu and Gd.  For neutrons of wavelength 4\AA, the respective $\delta/\beta$ ratios are $2.0 \times 10^5$, $8.7 \times 10^4$, $7.3 \times 10^3$, and $4.8 \times 10^{-1}$ \cite{Pfeiffer2009}.  Hence, relative to the maximal SNR boost that would be expected for Pb at the stated wavelength, the expected SNR boosts for the four indicated elements are in the ratio $1:8\times 10^{-2}:5\times10^{-5}:1\times10^{-17}$.  Thus, for a fixed sample size and energy range, and out of the list of chosen elements and the chosen neutron energy, Pb shows the highest neutron phase contrast (and associated energy boost), with Si and Cu showing an intermediate level of phase contrast, and Gd showing extremely weak neutron phase contrast

\section{Experiment}\label{Sec:Experiment}

The proof-of-concept measurement was performed at IMAT (Imaging and Materials Science \& Engineering) neutron imaging and diffraction instrument at ISIS, Oxfordshire, United Kingdom \cite{IMAT1,IMAT2,IMAT3}.  This pulsed spallation source yields both thermal and cold neutrons, with 50 kW power at 10 Hz \cite{IMAT1}.  Downstream of both source and moderator lie a neutron guide, wavelength-band choppers, beam-shaping elements (pinhole plus slits), sample stage and position-sensitive Multi-Channel-Plate (MCP) Timepix 2 detector \cite{IMAT2}. Energy resolution is obtained using time of flight analysis \cite{IMAT1,IMAT2}.  Exposure times were $\sim$6~minutes per tomographic projection, binned into 524 equal-width wavelength bins ranging from 0.70~\AA{} to 6.76~\AA.  The pinhole-to-sample distance was $L=10$~m, and the pinhole diameter was $d=40$~mm. The divergence was $\Theta=\tan^{-1}(d/L)=0.004$~radians.  201 tomographic projection angles were used, ranging between $0^{\circ}$ and $360^{\circ}$ in equal angular steps of  $1.8^{\circ}$.  2D projection images were taken on a 256$\times$256 pixel array, corresponding to one quarter of the available detector surface.  The pixel size was $55\times55~\mu$m$^2$, and the sample-to-detector distance was $\Delta\approx30$~mm.  The sample was assumed to be composed of amorphous $^{12}{\textrm{C}}$, giving a scattering length $b=6.65\times10^{-15}$~m and total neutron cross section $\sigma=5.55\times10^{-28}$~m$^2$.







\begin{figure}
    \centering
    \includegraphics[width=0.75\columnwidth]{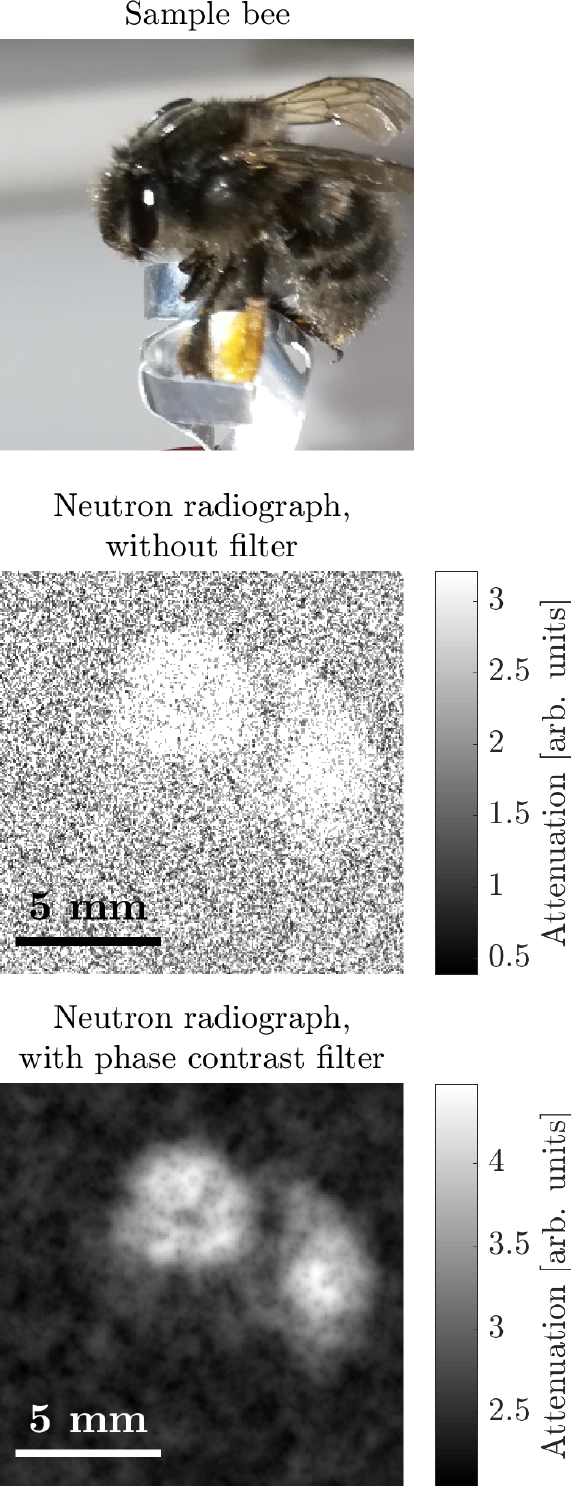}
    \caption{{\it Top}: Image of bee used for measurements. {\it Middle}: Neutron propagation-based phase-contrast image data, for one projection and for $\lambda = 5.919 \pm  0.006$ \AA, without phase contrast filtering. {\it Bottom}: Corresponding neutron attenuation data with phase contrast filter applied. }
    \label{fig:Bee}
\end{figure}

The sample was a common honey bee, shown in the upper panel of Fig.~\ref{fig:Bee}. The middle panel of the same figure shows a sample 2D projection image, obtained using the experimental parameters listed above.  Here, the selected neutron wavelength was $\lambda = 5.919$~\AA{} $\pm~0.006$~\AA, corresponding to one of the 524 wavelength bins. The event rate recorded by the utilized part of the detector in the selected wavelength bin was 110~neutrons/s/cm$^2$, this being 0.07\% of events in the full wavelength range. This event rate, together with the stated exposure time and pixel dimensions, corresponds to an average of 1.2 neutrons per pixel, hence the SNR of the detected tomographic projection is on the order of unity. 

Before proceeding, we note that (i) this detected event rate of one neutron per pixel in each tomographic projection has been chosen to demonstrate the ultimate limits of the method \cite{Goy2018}, rather than to obtain images that can compete with the current state of the art; (ii) Paganin-type phase retrieval has previously been successfully applied to quantitative analyses using transmission electron microscope images with on the order of one detected electron per pixel \cite{Clark2019}, which is consistent with the suggestion that one imaging quantum per pixel represents the ultimate threshold for the minimum radiant exposure that can be accommodated by the method (cf.~\citet{Goy2018} and \citet{Johnson2020}).  

Application of the phase-retrieval decoding based on Eqs.~(\ref{eq:PaganinBeltranMethodForNeutrons}) and (\ref{eq:tau}), to the tomographic projection in the middle panel of Fig.~\ref{fig:Bee}, yielded the image in the lower panel of Fig.~\ref{fig:Bee}.  A significant boost in SNR, and therefore in the effective degree of brilliance that may be associated with this propagation-based phase-contrast imaging scenario,  is clearly evident in passing from the middle panel to the lower panel of Fig.~\ref{fig:Bee}.  This SNR boost is consistent with similar observations made in studies using both x rays \cite{SNRboost1,SNRboost2,SNRboost3,SNRboost4,SNRboost5} and electrons \cite{Clark2019}.  In information-theoretic terms \cite{MacKay2003}, the positive effect of the phase-contrast-induced signal-encoding and subsequent phase-retrieval decoding, when offset with the negative effect of the increased degree of collimation needed to make the phase contrast operative, has led to a net increase in the SNR of the decoded signal \cite{gureyev2017unreasonable}.  

\begin{figure}
    \centering
    \includegraphics[width=\columnwidth]{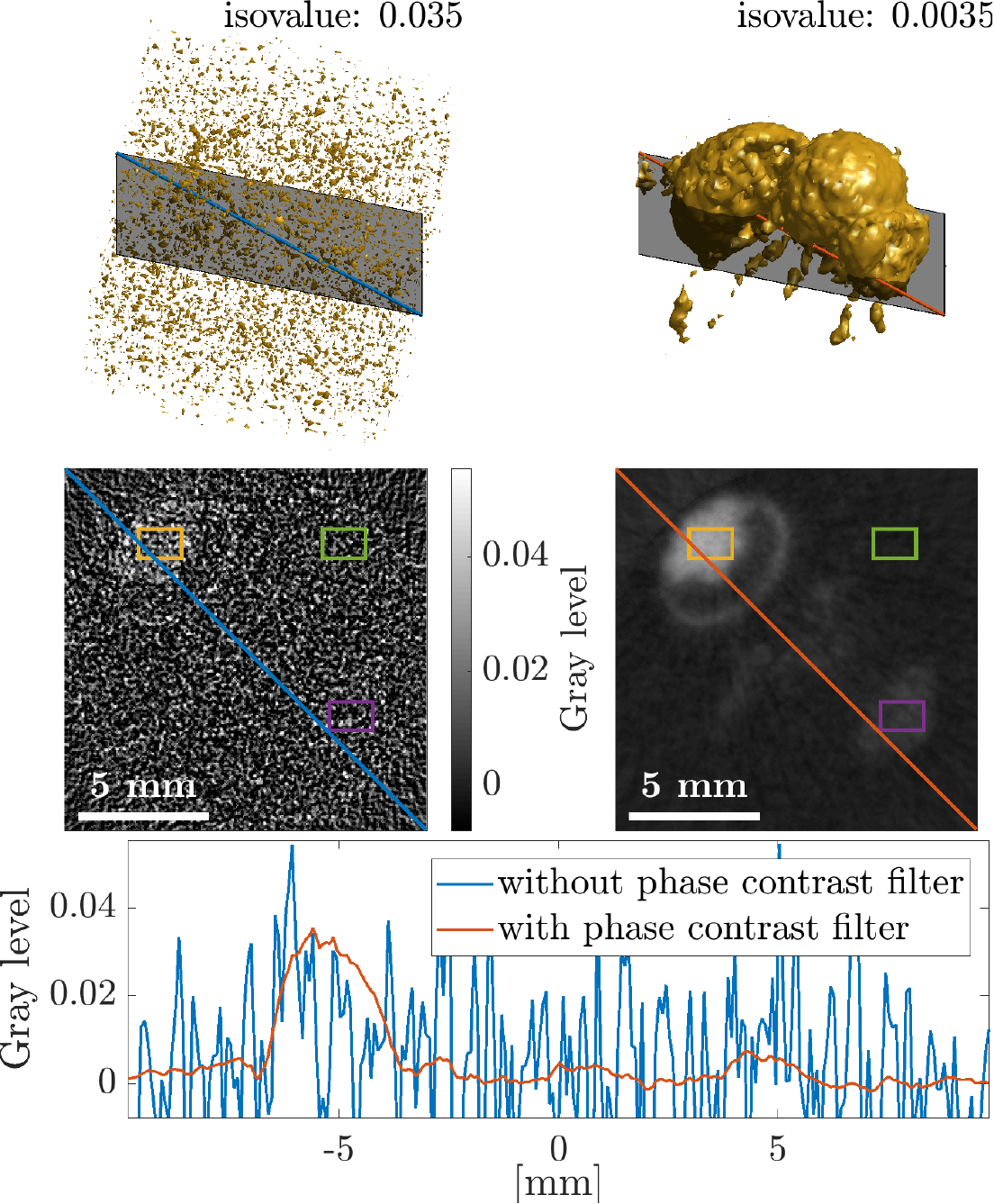}
    \caption{3D tomographic reconstruction of measured bee. Left column and blue curve are from parallel backprojection of data without the phase contrast filter. Right column and orange curve are with phase contrast filtering applied to recorded radiographs. The planes drawn in the top images indicate the slice shown in the middle row. The lines indicate the path along which the reconstruction is shown in the bottom plot. The purple, yellow, and green boxes indicate regions of interest selected for the calculation of SNR boosts at the end of Sec.~\ref{Sec:Experiment}. For an animation showing all slices, see the movie contained in the Supplementary Material.}
    \label{fig:recon}
\end{figure}

The results of the subsequent tomographic-reconstruction step, based on filtered back-projection \cite{KakSlaneyBook}, are shown in Fig.~\ref{fig:recon}.  The top-left (isosurface rendering) and middle-left (tomographic slice of recovered density) panels correspond to filtered-backprojection tomographic reconstruction \cite{KakSlaneyBook} being applied to the raw projection data.  The corresponding post-phase retrieval data yields the iso-surface rendering and tomographic-slice images shown in the top-right and middle-right of Fig.~\ref{fig:recon}, respectively.  The bottom row of Fig.~\ref{fig:recon} shows line profiles of the reconstructed tomographic density, with the blue gray-level profile corresponding to the diagonal trace in the middle-left panel (i.e.,~without the phase-retrieval step) and the orange gray-level profile corresponding to the diagonal trace in the middle-right panel (i.e.,~with the phase retrieval step).      

{\em Validity condition for the transport-of-intensity equation:} Using similar triangles applied to the geometry in Fig.~\ref{fig:SetUp}, together with the numerical values listed above, the resolution $R$ associated with penumbral blurring can be crudely estimated as:
\begin{equation}\label{eq:SR}
R=\frac{d\Delta}{L}\approx 0.1~{\textrm{mm}}.   
\end{equation}
This resolution of 100 $\mu$m is well-matched to the pixel size of 55 $\mu$m.  It corresponds to Fresnel number $N_{\textrm{F}}$ \cite{SalehTeichBook} 
\begin{equation}\label{eq:FresnelNumber}
    N_{\textrm{F}}=\frac{R^2}{\lambda \Delta}\approx 800.
\end{equation}
This Fresnel number easily satisfies the validity condition 
\begin{equation}\label{eq:FresnelNumberValidity}
    N_{\textrm{F}} \gg 1
\end{equation}
for the transport-of-intensity equation \cite{teague1983}, upon which our analysis is based (see Eq.~(\ref{eq:TIE})).

{\em Collimation condition:}  With the specified values of $b, \lambda,\sigma,\Delta$ as listed above, the collimation condition ``rule of thumb'' in  Eq.~(\ref{eq:CollimationCondition2}) gives the critical divergence (minimum degree of collimation) as $\Theta_{\textrm{critical}}=75^{-1}=0.013$ radians.  The actual divergence of $\Theta=250^{-1}=0.004$ radians therefore meets the collimation condition in Eq.~(\ref{eq:CollimationCondition2}).  Also, from the simple ``rule of thumb'' in  Eq.~(\ref{eq:OptimumDivergence}), the optimum divergence is $\Theta_{\textrm{optimum}}=125^{-1}=0.008$ radians. Note that the utilized divergence is double that which would be routinely used for attenuation-based neutron imaging (i.e.,~$\Theta_0\approx 125^{-1}$~radians \cite{IMAT1,IMAT2}), hence there is a four-fold reduction in flux that is implied by collimating the beam in this experiment, to improve the spatial coherence for the purpose of increasing the propagation-based phase contrast.

{\em Condition for boosting of the effective degree of brilliance:} With the specified numerical values for $b,\lambda,\sigma,\Delta$, the condition in Eq.~(\ref{eq:Y}) for effective-brilliance boosting becomes $\Theta_0<32$ radians, which is always true; recall that $\Theta_0\approx125^{-1}$~radians.  This establishes the experiment to indeed be in an SNR-boosting regime, a fact which is otherwise clear from the experimental reconstructions in Figs~\ref{fig:Bee} and \ref{fig:recon}.  
To estimate the actual SNR boost obtained in this experiment, we use the purple and green boxed regions in the left-middle row of Fig.~\ref{fig:recon} to estimate the pre-phase-retrieval SNR, and use the purple and green boxed regions in the right-middle row of Fig.~\ref{fig:recon} to estimate the post-phase-retrieval SNR.  The ratio of these SNRs gives an SNR boost, due to the phase-retrieval step, of $45\pm 1$ (error bars estimated using similar regions of interest).  If we instead use the yellow and green boxed regions, corresponding to a part of the object that has significantly smaller density, the estimated SNR boost (for the phase-retrieval step) is also $45\pm 1$.   The SNR boost is relative to the previously-mentioned collimation-related flux reduction of a factor of 4, hence the effective SNR boost needs to be divided by $\sqrt{4}=2$ in order to properly quantify the {\em net} effect of (i) SNR reduction due to sufficiently-hard collimation needed to achieve propagation-based phase contrast, followed by (ii) SNR increase due to the subsequent phase-retrieval step. Thus the measured {\em net} SNR boost is $23\pm1$, corresponding to a net effective-brilliance boost of $530 \pm50$.  Our experiments are therefore consistent with a boost in effective neutron brilliance of over two orders of magnitude.  This may be compared to the current state-of-the-art in x-ray experiments, which have reported SNR boosts well in excess of four orders of magnitude using Paganin-type phase retrieval \cite{SNRboost1,SNRboost2,SNRboost3,SNRboost4,SNRboost5}.  However, we again emphasize that the situation is not as favorable for neutrons as it is for x rays: (i) the boost in the effective degree of brilliance is only two orders of magnitude in our neutron experiment, rather than over four orders of magnitude in contemporary x-ray experiments, and (ii) we have a more restricted domain of validity for the case of neutrons, when compared to the case of x rays.


\section{Discussion}\label{Sec:Discussion}

Numerous public-domain software implementations exist, for the x-ray version of Eq.~(\ref{eq:PaganinBeltranMethodForNeutrons}) \cite{paganin2002}. These include ANKAphase \cite{ANKAphase}, X-TRACT \cite{XTRACT}, pyNX \cite{FavreNicolin2011}, PITRE \cite{Chen2012}, Octopus \cite{Boone2012,Dierick2004}, pyHST2 \cite{pyHST2}, TomoPy \cite{TomoPy}, SYRMEP TomoProject \cite{Brun2017}, pyXIT \cite{pyXIT}, HoloTomo Toolbox \cite{Lohse2020}, PyPhase \cite{PyPhase2020}, Livermore Tomography Tools \cite{CHAMPLEY2022102595}, Tofu \cite{Farago:tv5034} and TomocuPy \cite{TomocuPy-arxiv}.  Since the x-ray and neutron versions are mathematically identical single-parameter reconstructions, the already-available software packages cited above---many of which also enable tomographic processing, either directly or by interfacing with standard tomographic software packages---may be utilized for propagation-based phase contrast neutron data without any modification.

Parallel $z$-directed neutrons have been assumed throughout the development of this paper.  However, all of the results readily  carry over to the case of divergent illumination from a small neutron source, on account of the Fresnel scaling theorem \cite{Pogany1997,paganin2002,paganin2006}.  Here, we simply replace the sample-to-detector propagation distance $\Delta$ by the scaled propagation distance $\Delta/M$, where $M$ is the geometric magnification associated with the point-projection geometry. Thus divergent-beam tomographic reconstructions may be performed, in both mono-energetic \cite{GureyevAPL2006} and poly-energetic settings \cite{Myers2007}. 

We conjecture that the assumption of a single material may be somewhat less restrictive than it might seem, particularly in three spatial dimensions (i.e.,~in a neutron-tomography context), where many samples of interest may be {\em locally} described as composed of a single material.  This suggestion, which would form an interesting avenue for further work, arises on account of previous x-ray studies. As shown by \citeauthor{SNRboost1}~\cite{SNRboost1,SNRboost2} for x rays, we can typically choose the parameter $\tau$ (see Eq.~(\ref{eq:tau})) corresponding to a material of interest, which will only locally corrupt the tomographic reconstruction of features composed of other materials. This locality arises from the fact that the real-space version of the phase-retrieval filter \cite{Ullher2015,Thompson2019}, which may be applied after rather than before tomographic reconstruction if the sample is weakly attenuating, has the same functional form as the manifestly-local Yukawa potential \cite{Yukawa1935}. Separate x-ray tomographic reconstructions may thus be performed for each material of interest, before splicing the resulting reconstructions together using the method of \citeauthor{SNRboost1} \cite{SNRboost1,SNRboost2}.  This strategy has proved useful for x-ray tomography and it would be interesting to investigate to what extent, if any, this extension to non-homogeneous objects might be possible for neutron phase-contrast tomography.  

What is the physical reason underpinning the significant SNR-boosting properties of Eq.~(\ref{eq:PaganinBeltranMethodForNeutrons})
?  We refer the reader to \citeauthor{gureyev2017unreasonable}~\cite{gureyev2017unreasonable} for the theoretical details, together with the previously mentioned information-theoretic principle that SNR may be boosted using a code--decode strategy when transmitting a signal through a noisy channel \cite{MacKay2003}, and here give a heuristic explanation to both motivate and augment the rigorous results developed in the just-cited references.  As pointed out earlier, the propagation-based neutron phase contrast image in Eq.~(\ref{eq:TIE-HomWithBlurring}) may be viewed as creating a sharpened version of the attenuation-based contact neutron image $I_0 \exp\left[-\sigma\rho_{\perp}({\bf r}_{\perp})\right]$.  This sharpening follows from the fact that this equation corresponds exactly to approximate deconvolution (sharpening); this has an associated transverse length scale $\ell$ given by the square root of the coefficient of the transverse Laplacian \cite{Unsharp0} in Eq.~(\ref{eq:TIE-HomWithBlurring}): 
\begin{align}
\ell=\sqrt{\frac{b\lambda^2\Delta}{2\pi\sigma}-\frac{\mathcal{A}}{8}}.
\end{align}
Indeed, from another albeit closely related perspective, Eq.~(\ref{eq:TIE-HomWithBlurring}) is mathematically identical in form to unsharp-mask image sharpening using a Laplacian kernel \cite{Unsharp0,Unsharp1,Unsharp2}.  This image sharpening is evident from the edge enhancement due to Fresnel diffraction.  Regardless of how we understand the image sharpening, the crucial point to note is that this sharpening occurs before the addition of noise in the detection process.  Moreover, if the aforementioned noise is white then it will be evenly spread through Fourier space, in contrast to the object which will typically have a Fourier-space power spectrum that decreases rapidly with increasing radial spatial frequency.  The Lorentzian Fourier-space filter in Eq.~(\ref{eq:PaganinBeltranMethodForNeutrons})
, which is a low pass filter since its function is to negate the previously mentioned sharpening, will therefore suppress noise much more strongly than it suppresses signal due to the object.  The result is a strong increase in SNR, corresponding to a boost in effective brilliance. 

As previously mentioned, the boost in the effective degree of brilliance may be used to reduce exposure times, for propagation-based neutron phase-contrast tomography and radiography, relative to its attenuation-based counterparts.  Alternatively, the SNR boost associated with the effective-brilliance increase may be traded off against increased contrast-to-noise ratio (CNR) and/or resolution, while keeping acquisition times relatively fixed.  Increased CNR follows directly from increased SNR.  Increased resolution follows from the usual trade-off between noise and resolution \cite{GureyevNRU,GureyevNRU2020}, so that e.g.~(i) an increase in SNR by a factor of $K$ but with no change in spatial resolution may be exchanged for (ii) no SNR increase but an increase in spatial resolution by a factor of $K$ (by using a smaller pinhole, whose area is $K \times K=K^2$ times smaller, than the pinhole that was previously used). 



Dark-field imaging, namely the imaging of scattering contrast due to unresolved micro-structure within a sample, is an important topic that has been only cursorily mentioned in this paper.  Indeed, the development of dark-field neutron tomography using a grating-based setup, by \citeauthor{Strobl2008}~\cite{Strobl2008}, is a particularly significant advance that is attracting much attention within the neutron-imaging community.  The propagation-based method for neutron phase-contrast imaging, which we consider in this paper, might be generalized to include dark-field/scattering contrast. This might be achieved using a formalism for paraxial imaging \cite{MorganPaganin2019,PaganinMorgan2019} based on the Fokker--Planck equation \cite{Risken1989,mandel1995optical}. This generalizes the transport-of-intensity equation \cite{teague1983}, upon which the present paper is based, enabling it to take into account local ultra-small-angle scattering and its associated diffusive transport, which augments the coherent transport associated with attenuation and refraction. To second  order in the sample-to-detector propagation distance $\Delta$, this forward-Kolmogorov Fokker--Planck generalization of the transport-of-intensity expression in Eq.~(\ref{eq:TIE}) is  \cite{MorganPaganin2019,PaganinMorgan2019}: 
 \begin{align}
 \label{eq:FokkerPlanckEquation}
\nonumber I({\bf r}_{\perp}, & z=\Delta) \approx  I({\bf r}_{\perp},z=0) \\ \nonumber
&-\frac{\lambda \Delta}{2 \pi}\nabla_{\perp}\cdot{[I({\bf r}_{\perp},z=0)\nabla_{\perp}\varphi({\bf r}_{\perp},z=0)]}
\\ 
\nonumber &+ \Delta^2\frac{\partial^2}{\partial x^2}\left[D^{(xx)}_{\textrm{eff}}({\bf{r}}_{\perp})I({\bf{r}}_{\perp},z=0) \right]  
  \\ \nonumber &+ \Delta^2\frac{\partial^2}{\partial y^2}\left[D^{(yy)}_{\textrm{eff}}({\bf{r}}_{\perp})I({\bf{r}}_{\perp},z=0)\right]  
  \\  &+ \Delta^2\frac{\partial^2}{\partial x\partial y}\left[D^{(xy)}_{\textrm{eff}}({\bf{r}}_{\perp})I({\bf{r}}_{\perp},z=0)\right].
\end{align}
Here, $D^{(xx)}_{\textrm{eff}}({\bf{r}}_{\perp})$, $D^{(yy)}_{\textrm{eff}}({\bf{r}}_{\perp})$ and $D^{(xy)}_{\textrm{eff}}({\bf{r}}_{\perp})$ are the components of an index-symmetric second-rank effective-diffusion tensor field that may be used to describe the position-dependent blur associated with locally-elliptical ultra-small-angle neutron scattering fans emerging from each point over the exit surface of the sample.  This tensor field quantifies the directional-dark-field signal associated with spatially-unresolved structure within the volume of the sample \cite{jensen2010a,jensen2010b,jud2017}, in the context of propagation-based neutron phase contrast, which bifurcates the neutron flow downstream of the sample into coherent and diffuse probability-current-density channels.   In particular, for the special rotationally-symmetric case where this second-rank diffusion-tensor field may be approximated as being proportional to the unit tensor, albeit with a constant of proportionality that depends on transverse position, the single-image single-material propagation-based phase-retrieval method of \citeauthor{paganin2002} \cite{paganin2002} may be generalized to a two-image method that is able to recover both the projected density and the projected ultra-small-angle-scattering signal of a single-material object \cite{PaganinMorgan2019,Leatham2021}, given two images taken at two different propagation distances.  Further exploration of the Fokker--Planck extension to the neutron phase-contrast imaging work of this paper, together with its additional extension using the paraxial-imaging Kramers--Moyal equation and its associated hierarchy of diffusion tensors \cite{MorganPaganin2019,PaganinMorgan2019} which are related to the various moments \cite{Modregger2017,Modregger2018} of the position-dependent ultra-small-angle neutron scattering fans, would be an interesting topic for future exploration.

Another interesting avenue, for investigation subsequent to the present paper, is to apply the polychromatic version of our algorithm, as developed in Eqs.~(\ref{eq:Polychromatic5}) and (\ref{eq:Polychromatic6}) from Appendix A, to non-energy-binned broad-band polychromatic neutron propagation-based phase contrast images obtained with smaller pinhole sizes.  See Appendix A for further details.  The idea of phase retrieval using broad-band polychromatic neutron phase contrast images has been considered previously, using a phase-retrieval method that amplifies rather than suppresses noise \cite{McMahon2003}.  However, the idea warrants revisiting in the context of the present paper, due to the SNR-boosting properties of Eqs.~(\ref{eq:Polychromatic5}) and (\ref{eq:Polychromatic6}) from Appendix A \cite{SNRboost1,SNRboost2,SNRboost3,SNRboost4,SNRboost5}.  For example, in the present study, time-of-flight monochromatization was achieved using 524 equally-spaced energy bins over the wavelength range from 0.70~\AA{} to 6.76~\AA.  If the pinhole diameter were to have been reduced from $d=40$~mm to $d=10$~mm, the resulting net-flux reduction of $4^2=16$ could easily have been compensated for by a net flux-boost by a factor of 1400,\footnote{For the experiment using energy-filtered neutrons reported in Sec.~III, the neutron flux recorded by the detector without the sample in the beam, over the utilized part of the detector at the selected wavelength bin of 5.9 \AA, was 110 neutrons/s/cm$^2$ (averaged over the acquisition time of 6 minutes).  For the entire wavelength range measured, the flux was about $1.56\times 10^5$ neutrons/s/cm$^2$ measured by the detector.  Hence the net boost in utilized flux, that is to be obtained by not needing any energy filter, is $1.56\times 10^5/110 \approx 1400$.} due to not needing to monochromatize and instead having a polychromatic image energy-integrating over all 524 energy bins. Indeed, by using fully poly-energetic neutrons, we anticipate that there would be a net increase in the utilized flux by a multiplicative factor of 1400/16=88, even with the smaller pinhole indicated above; this would boost the SNR of the raw phase-contrast data by a factor of $\sqrt{88}\approx9$, relative to that utilized in the present paper, with an additional boost in SNR to be expected on account of the significantly improved spatial coherence that would be due to the use of a smaller pinhole (cf.~\citeauthor{Wilkins1996} \cite{Wilkins1996} for an early example of this last-mentioned fact, in the x-ray literature). Also, the corresponding Fresnel number would have been reduced from 800 (see Eq.~(\ref{eq:FresnelNumber})) to $N_{\textrm{F}}\approx 800/4^2=50$, which is still well within the validity condition in Eq.~(\ref{eq:FresnelNumberValidity}).  Lastly, the spatial resolution in Eq.~(\ref{eq:SR}) would become 25~$\mu$m, which remains well-matched to the pixel size of 55~$\mu$m.    

This study has been restricted to positive defocus $\Delta$, since free-space propagation was utilized as the mechanism for phase contrast. However, as briefly mentioned earlier in the paper, negative defoci will be accessible if an imaging system is interposed between the sample and the detector, e.g.~in possible future applications of the method of this paper to neutron microscopy \cite{NeutronMicroscope1,NeutronMicroscope2} using compound refractive lenses \cite{NeutronCRL,NeutronCRL2}.  For such systems, the requirement for $b\Delta$ to be positive, so as to avoid a division-by-zero divergence in the Fourier filter given in Eq.~(\ref{eq:PaganinBeltranMethodForNeutrons}), implies that either (i) the scattering length $b$ for the material in the single-material sample, and the defocus $\Delta$, should both be positive; or (ii) $b$ and $\Delta$ should both be negative.  In the context of neutron imaging systems, for which residual optical aberrations may be present, we note that \citeauthor{Liu2011a} have developed a form of the Paganin algorithm which takes both defocus and spherical aberration into account \cite{Liu2011a,Liu2011b}.


\section{Conclusion}\label{Sec:Conclusion}

In light of the possibilities for a boost in the effective degree of brilliance that may be available for propagation-based phase-contrast neutron tomography, we believe it is timely that this modality for three-dimensional imaging be revisited.  Theoretical and experimental evidence was presented to support this suggestion, for the limited class of samples that can be approximated as being composed of a single material.  Within its rather restricted domain of applicability, the proof-of-concept experiment implementing the method was able to achieve boosts in the effective degree of brilliance of greater than two orders of magnitude. This boost may be used to reduce exposure times, or instead one may want to keep longer exposure times and harness the effective-brilliance boost to improve contrast.

\acknowledgements{We thank the Danish Agency for Science, Technology, and Innovation for funding the instrument center DanScatt. We acknowledge useful discussions with Joseph Bevitt, Jeremy Brown, Laura Clark, Linda Croton, Carsten Detlefs, Margaret Elcombe, Ulf Garbe, Timur Gureyev, Andrew Kingston, Luise Theil Kuhn, Kieran Larkin, Kaye Morgan, Thomas Leatham, Glenn Myers, Daniele Pelliccia, Tim Petersen, Kirrily Rule, and Floriana Salvemini.}

\appendix

\section{Application to poly-energetic neutron beams}

Poly-energetic neutron beams are able to yield propagation-based phase contrast \cite{McMahon2003,Jacobson2004} (cf.~analogous work in the x-ray regime \cite{Wilkins1996}).  We now use a similar argument to that of several x-ray papers on phase retrieval using poly-energetic radiation \cite{Gureyev2006,Myers2007,Arhatari2008}, to suggest how the method of Eq.~(\ref{eq:PaganinBeltranMethodForNeutrons}) may be applied to poly-energetic neutron beams, for the case of weakly-attenuating samples that are composed of a single material.  

The assumption of weak attenuation allows us to make the approximation
\begin{equation}
    \exp\left[-\sigma\rho_{\perp}({\bf r}_{\perp})\right] \approx 1 -\sigma\rho_{\perp}({\bf r}_{\perp})
\end{equation}
in the final line of Eq.~(\ref{eq:TIE-HomWithBlurring}), leaving us with: 
\begin{align}\label{eq:Polychromatic}
\frac{I_E({\bf r}_{\perp},z =\Delta)}{I_{0,E}} &= \left[ 1-\tau_E\nabla_{\perp}^2\right] \left[1-\sigma_E\rho_{\perp}({\bf r}_{\perp})\right] \\ \nonumber &=1-\sigma_E\rho_{\perp}({\bf r}_{\perp})+\sigma_E\tau_E\nabla_{\perp}^2\rho_{\perp}({\bf r}_{\perp}).
\end{align}
In the above equation, we have (i) made use of the definition for $\tau$ in Eq.~(\ref{eq:tau}), and (ii) explicitly indicated the energy dependence of various quantities via a subscript $E$.  

If we multiply through by the neutron energy spectrum $I_{0,E}$ and then average over energies, as indicated via an overline, we obtain:
\begin{align}\label{eq:Polychromatic2}
\overline{I_E({\bf r}_{\perp},z =\Delta)} =\overline{I_{0,E}}-(\overline{I_{0,E}\sigma_E}-\overline{I_{0,E}\sigma_E\tau_E}\nabla_{\perp}^2)\rho_{\perp}({\bf r}_{\perp}).
\end{align}

Upon introducing the spectrally-averaged quantities defined via
\begin{subequations}
\begin{eqnarray}\label{eq:Polychromatic3}
    I_{\textrm{av}}({\bf r}_{\perp},z =\Delta) &\equiv& \overline{I_E({\bf r}_{\perp},z =\Delta)}/\overline{I_{0,E}}, \\ \sigma_{\textrm{av}} &\equiv& \overline{\sigma_E I_{0,E}}/\overline{I_{0,E}},\\ (\sigma\tau)_{\textrm{av}} &\equiv& \overline{\sigma_E \tau_E I_{0,E}}/\overline{I_{0,E}},
\end{eqnarray}
\end{subequations}
and applying the weak-attenuation approximation in reverse (this is analogous to passing from the second line to the first line of Eq.~(\ref{eq:Polychromatic})), we obtain the following poly-energetic variant of Eq.~(\ref{eq:TIE-HomWithBlurring}):
\begin{align}\label{eq:Polychromatic4}
I_{\textrm{av}}({\bf r}_{\perp},z=\Delta)=\left[1-\frac{(\sigma\tau)_{\textrm{av}}}{\sigma_{\textrm{av}}}\nabla_{\perp}^2\right]\exp[-\sigma_{\textrm{av}}\rho_{\perp}({\bf r}_{\perp})].
\end{align}

Solving for the projected density, we obtain a poly-energetic version of Eq.~(\ref{eq:PaganinBeltranMethodForNeutrons}) that is valid for weakly-attenuating single-material samples:
\begin{align}\label{eq:Polychromatic5}
\rho_{\perp}({\bf r}_{\perp})=   -\frac{1}{\sigma_{\textrm{av}}}\log_{\textnormal{e}}\left(\mathcal{F}^{-1}\left\{\frac{\mathcal{F}[I_{\textrm{av}}({\bf r}_{\perp},z=\Delta)]}{1+\frac{(\sigma\tau)_{\textrm{av}}}{\sigma_{\textrm{av}}}(k_x^2+k_y^2)}\right\}\right).
\end{align}

This is mathematically identical in form to the mono-energetic version of the algorithm in Eq.~(\ref{eq:PaganinBeltranMethodForNeutrons}), aside from the replacements:
\begin{equation}\label{eq:Polychromatic6}
 \sigma\longrightarrow \sigma_{\textrm{av}},\quad \tau\longrightarrow \frac{(\sigma\tau)_{\textrm{av}}}{\sigma_{\textrm{av}}},    
\end{equation}
together with the use of a poly-energetic neutron phase contrast image rather than an energy-filtered image.  

This poly-energetic form may be advantageous, since there will be a significant boost in utilizable neutron flux on account of there being no need to energy-filter broad-band poly-energetic neutron phase-contrast images.  However, we must emphasize that the actual experimental evidence in the main text of our paper relates to the monochromatic case, with the experimental testing of the above polychromatic generalization to the theory being left to a subsequent paper \cite{Ostergaard2022polychromatic}.

\section{Visibility boost induced by free-space propagation}

Here we justify the statement, made in Sec.~\ref{Sec:ChoiceOfCollimation}, that the term in large round brackets in the final line of Eq.~(\ref{eq:TIE-HomWithBlurring}), namely $\tau$ as given in Eq.~(\ref{eq:tau}), is a measure of the visibility of the propagation-based phase contrast features that we  would observe in the high-neutron-flux limit (zero noise limit).  

Work with one transverse dimension $x$, for simplicity.  Consider a weakly attenuating single-material sample whose projected number density has the form of a sinusoidal grating:
\begin{equation}\label{eq:SinusoidalGrating}
    \rho_{\perp}(x)=\rho_0 [\sin(2 \pi x/p)+1].
\end{equation}
Here, $\rho_0$ is a positive constant, and $p$ denotes the transverse period of the grating.

The attenuation-contrast image $I_{\textrm{abs}}(x,z=0)$ will have the intensity profile given by Eq.~(\ref{eq:IntensityProjectionApproximation}) as
\begin{equation}
  I_{\textrm{abs}}(x,z=0)=I_0 \exp\{-\sigma \rho_0   [\sin(2 \pi x/p)+1]\}.    
\end{equation}
The corresponding Michelson visibility \cite{Michelson1927Book} is
\begin{eqnarray}
    \nonumber \mathcal{V}_{\textrm{abs}} &=& \frac{\max[I_{\textrm{abs}}(x,z=0)]-\min[I_{\textrm{abs}}(x,z=0)]}{\max[I_{\textrm{abs}}(x,z=0)]+\min[I_{\textrm{abs}}(x,z=0)]} \\ \nonumber &=& 
    \frac{I_0-I_0\exp(-2\sigma\rho_0)}{I_0+I_0\exp(-2\sigma\rho_0)} \\ &\approx& \sigma\rho_0,
\end{eqnarray}
where ``max'' and ``min'' respectively denote the maximum or minimum value of the corresponding argument, and the assumption of weak attenuation has been used to make the approximation
\begin{equation}
\exp(-2\sigma\rho_0)\approx 1 -2\sigma\rho_0.
\end{equation}

For the propagation-based phase contrast image $I_{\textrm{prop}}(x,z=\Delta)$, Eqs.~(\ref{eq:TIE-HomWithBlurring}), (\ref{eq:tau}) and (\ref{eq:SinusoidalGrating}) give, in the weak-attenuation limit,
\begin{eqnarray}
     \frac{I_{\textrm{prop}}(x,z=\Delta)}{I_0}  \quad\quad\quad\quad\quad\quad\quad\quad\quad\quad\quad\quad\quad\quad\quad \\ \nonumber  =1-\sigma\rho_0-\sigma\rho_0\left(1+\frac{4\pi^2\tau}{p^2} \right)\sin\left(  \frac{2\pi x}{p}\right).
\end{eqnarray}
The corresponding Michelson visibility is
\begin{equation}
    \mathcal{V}_{\textrm{prop}}=\rho_0\sigma\left(1+\frac{4\pi^2 \tau}{p^2}\right).
\end{equation}
This displays the linear proportionality with $\tau$ that we set out to demonstrate. 

We close this second appendix by noting that the boost in visibility, obtained in using propagation-based phase contrast in comparison to attenuation contrast, is also directly proportional to $\tau$:
\begin{equation}
    \frac{\mathcal{V}_{\textrm{prop}}}{\mathcal{V}_{\textrm{abs}}}=1+\frac{4\pi^2 \tau}{p^2}.
\end{equation}
This visibility boost is the signal-encoding process, induced by the diffraction physics of free-space propagation between the exit surface of the sample and the entrance surface of the position-sensitive detector, that underpins the possibilities explored in this paper.

\bibliography{refbibtek}

\end{document}